\documentstyle[12pt]{article}

\begin{document}

\hfill UB-ECM-PF 94/26

\vspace*{4mm}

\begin{center}
{\Large\bf
Manifestations of Quantum Gravity \\
in Scalar QED Phenomena }\\

\vspace{1cm}

\renewcommand{\thefootnote}{1}

{\bf Emilio Elizalde}
\footnote{E-mail: eli@zeta.ecm.ub.es. Address june-september 1994:
Department of Physics, Faculty of Sciences, Hiroshima University,
Higashi-Hiroshima 724, Japan.
E-mail: elizalde@aso.sci.hiroshima-u.ac.jp.}
\\
Center of Advanced Studies, CSIC,
Cam\'{\i} de Sta. B\`arbara, 17300 Blanes, \\
Department E.C.M. and I.F.A.E.,
Faculty of Physics, University of  Barcelona, \\
Diagonal 647, 08028 Barcelona,
Catalonia,
Spain \\

\renewcommand{\thefootnote}{2}

\vskip 0.5truecm
{\bf Sergei D. Odintsov}
\footnote{On leave of absence from Tomsk Pedagogical Institute,
634041 Tomsk, Russian Federation. E-mail: odintsov@ebubecm1.bitnet }
and {\bf August Romeo},
\\
Department E.C.M., Faculty of Physics,
University of  Barcelona, \\  Diagonal 647, 08028 Barcelona,
Catalonia,
Spain \\

\end{center}

\vspace{1.5cm}

\noindent{\large\bf Abstract.}

 Quantum gravitational corrections to the
effective potential, at one-loop level and in the leading-log
approximation, for scalar quantum electrodynamics with higher-derivative
 gravity ---which is taken
as an effective theory for quantum gravity (QG)--- are calculated.  We
point out the appearence of relevant
phenomena caused by quantum gravity, like dimensional transmutation,
 QG-driven instabilities of
the potential, QG corrrections to scalar-to-vector mass ratios, and
curvature-induced phase transitions, whose existence is shown by means of
analytical and numerical study.

\vspace{1cm}

\begin{center}
\end{center}

\newpage

As is known, Einstein gravity is non-renormalizable
\cite{DW} and, therefore, it cannot play the role of a fundamental
theory of quantum gravity (QG). In the
best case, it might be regarded as an effective
model of some consistent theory of QG, still unknown to us
(for a recent discussion see \cite{D}).
In principle, one is allowed to consider
 alternative candidates as effective QG models
(for an introduction to effective theories see \cite{Wetal}), as the
effective
theory for the conformal factor, whose aim is to describe QG in the
 far infrared limit
\cite{AM}, among others.

A natural idea is to add higher-derivative terms to the Einstein gravity
action. In this way we obtain a higher-derivative QG which may be
envisaged as an effective theory for QG on the same level as the
Einsteinian one. This theory is known to be multiplicatively
renormalizable \cite{S}, even in the presence of matter \cite{BS}
(for a review, see \cite{BOS}), and also asymptotically free. (Due to
its perturbative non-unitarity this theory cannot be considered
as an eligible candidate for a fundamental QG model but it is perfectly
valid as an effective theory).

Since higher-derivative QG contains not only dimensional couplings
---Newton's $16 \pi G$ and the cosmological $\Lambda$--- but also
dimensionless coupling constants, one could expect some qualitative
differences in its influence on physics near the
Planck-mass scale
($\mu_{\mbox{\scriptsize Pl}}\simeq 1.2 \cdot 10^{19}$ GeV)
or between the Planck-mass and the GUTs energy scales
($\mu_{\mbox{\scriptsize GUT}} \simeq 10^{15}$ GeV), as compared to the
Einstein theory. In particular, there appear QG radiative corrections
to the beta functions of the
 matter dimensionless coupling constants \cite{BS,BOS}.
It is precisely the purpose of the present letter to study QG
corrections
to one-loop and renormalization group (RG)
improved effective potentials (EP) of massless scalar
QED interacting with $R^2$-gravity. In this example we will show the
existence of a variety of interesting phenomena:
dimensional transmutation due to QG,
corrections to scalar-to-vector mass ratios,
QG-driven instabilities of EP,
and curvature-induced phase transitions.

Let us start from the multiplicatively renormalizable theory with the
Lagrangian \cite{BOS}
\begin{eqnarray}
{\cal L}&=&\displaystyle {1 \over \lambda}W
-{\omega \over 3 \lambda} R^2
+{1 \over 2} \xi R \varphi^2 \nonumber \\
&&+\displaystyle{1 \over 2}\left(
\partial_{\mu}\varphi_1 -eA_{\mu}\varphi_2
\right)^2
+{1 \over 2}\left(
\partial_{\mu}\varphi_2 -eA_{\mu}\varphi_1
\right)^2
-{1 \over 4!} f \varphi^4 -{1 \over 4}F_{\mu \nu} F^{\mu \nu},
\label{Lagr}
\end{eqnarray}
where $W=C_{\mu \nu \alpha \beta} C^{\mu \nu \alpha \beta}$
and $\varphi^2=\varphi_i \varphi_i$. The
$\lambda, \omega, \xi$ are gravitational
couplings, and the gravitational, scalar and vector fields are
quantized. One could also have started from
the massive model (including then also the Einstein theory) but,
since we are concerned with the approximation where the
background scalar satisfies $\phi^2 \gg m^2$, being $m^2$ the largest
effective mass of the theory, massive terms are not essential for
our purposes and it is enough to work with (\ref{Lagr}). Setting
$\omega=0, \xi=1/6$ in (\ref{Lagr}), we obtain conformally-invariant
gravity with scalar QED. Such a theory is multiplicatively
renormalizable only
when using the so-called special conformal regularization (for a
discussion and a list of references, see  \cite{BOS}).

Next, we find the effective potential \cite{CW,W} for (\ref{Lagr})
on a scalar-gravitational background. The result for the RG-improved
potential (for details on this formalism see
\cite{CW,EJ}, in  flat space, and \cite{EO}, in  curved space) can be
obtained as follows: using the linear curvature approximation and
$\varphi^2 \gg |R|$, i.e. expanding $V$ in powers of the curvature,
up to linear terms, we arrive at
\begin{equation}
V={1 \over 4! } f(t) \varphi^4(t)
-{1 \over 2} \xi(t) R \varphi^2(t),
\label{VRGi}
\end{equation}
where the system for the effective coupling constants may be found as
\[
\lambda(t)={ \lambda \over
\displaystyle 1+ { \alpha^2\lambda t \over (4 \pi)^2 } }, \hspace{1cm}
\alpha^2= {203 \over 15}, \hspace{1cm}
e^2(t)={ e^2 \over
\displaystyle 1- { 2 e^2\lambda t \over 3 (4 \pi)^2 } },
\]
\begin{equation}
\begin{array}{lll}
\displaystyle{d\omega \over dt}=\beta_{\omega}&=&
\displaystyle-{1 \over (4 \pi)^2}\lambda\left[
{10 \over 3}\omega^2 +\left( 5+\alpha^2 \right)\omega +{5 \over 12}
+3\left( \xi-{1 \over 6} \right)^2 \right], \\
\displaystyle{d\xi\over dt}=\beta_{\xi}&\equiv&
\beta_{\xi}^{(0)}+\Delta\beta_{\xi},\\
&&\displaystyle\beta_{\xi}^{(0)}={1 \over (4 \pi)^2}
\left( \xi -{1 \over 6} \right) \left( {4 \over 3}f -6e^2 \right), \\
&&\displaystyle\Delta\beta_{\xi}={1 \over (4 \pi)^2}\lambda\xi \left[
-{3 \over 2}\xi^2 +4\xi +3 +{10 \over 3}\omega
+{1 \over \omega}\left( -{9 \over 4}\xi^2 +5 \xi+ 1 \right)
\right], \\
\displaystyle{df \over dt}=\beta_f&\equiv&\displaystyle\beta_f^{(0)}
+\Delta\beta_f, \\
&&\displaystyle\beta_f^{(0)}={1 \over (4 \pi)^2}
\left( {10 \over 3}f^2 -12e^2f +36 e^4 \right), \\
&&\displaystyle\Delta\beta_f={1 \over (4 \pi)^2}
\left[
\lambda^2\xi^2\left( 15+ {3 \over 4\omega^2}
-{9 \xi \over \omega^2}+{27 \xi^2 \over \omega^2} \right) \right. \\
&&\hspace{6em}\displaystyle\left. -\lambda f\left(
5+3\xi^2+{33 \over 2\omega}\xi^2
-{6 \over \omega}\xi +{1 \over 2 \omega} \right) \right], \\
\displaystyle-{1 \over \varphi}{d\varphi\over dt}=
\gamma_{\varphi}&\equiv&
\displaystyle\gamma_{\varphi}^{(0)}+\Delta\gamma_{\varphi}, \\
&&\displaystyle\gamma_{\varphi}^{(0)}=-{3e^2 \over (4 \pi)^2}, \\
&&\displaystyle\Delta\gamma_{\varphi}={1 \over (4 \pi)^2}
{\lambda \over 4}\left(
{13 \over 3} -8\xi -3\xi^2 -{1 \over 6 \omega} -{2 \xi \over \omega}
+{3 \xi^2 \over 2 \omega}
\right), \\
\end{array}
\label{betafs}
\end{equation}
and $\displaystyle t={1 \over 2}\log {\varphi^2 \over \mu^2}$, with
the mass parameter $\mu^2$ taken to be of the order of
$\mu_{\mbox{\scriptsize Pl}}^2$.
The QG effective
coupling constants and QG corrections to the matter beta functions
have been
calculated in \cite{BS,BOS}. The Landau gauge is used for the vector
field and the harmonic gauge for the gravitational field. As one can see
from (\ref{VRGi}), (\ref{betafs}), the RG-improved EP is given in a
rather non-explicit form, but can be unveiled after some numerical work.

In the conformally invariant version of (\ref{Lagr}), the eqs. for
$\omega$ and $\xi$ disappear, and the only changing pieces of eqs.
(\ref{betafs}) are \cite{BOS,BS}
\[ \alpha^2 = {27 \over 2}, \]
\begin{equation}
\Delta\beta_f={1 \over (4 \pi)^2}
\left( {5 \over 12}\lambda^2 -{41 \over 8} \lambda f \right),
\hspace{1cm}
\Delta\gamma_{\varphi}={1 \over (4 \pi)^2}{27 \over 32} \lambda.
\label{betafsconf}
\end{equation}
Before going on to the numerical study of (\ref{VRGi}), let us consider
the much simpler one-loop EP, which can easily be obtained from
(\ref{VRGi}).
If we use Coleman-Weinberg normalization conditions \cite{CW}, it
reads
\begin{equation}
\begin{array}{lll}
V^{(1)}&=&\displaystyle{1 \over 4!} f\varphi^4
+{1 \over 48} (\beta_f -4f \gamma_{\varphi}) \varphi^4
\left( \log {\varphi^2 \over \mu^2 } - {25 \over 6} \right) \\
&&\displaystyle-{1 \over 2}\xi R \varphi^2
-{1 \over 4} (\beta_{\xi} -2\xi \gamma_{\varphi}) R \varphi^2
\left( \log {\varphi^2 \over \mu^2 } -3 \right),
\end{array}
\label{V1}
\end{equation}
where all the beta functions have already been defined in
(\ref{betafs}).
Expression (\ref{V1}), which is valid for any arbitrary massless
renormalizable theory, was first obtained in \cite{BO}. For its
conformal version, one just has to set $\xi=1/6$, $\beta_{\xi}=0$ and
take the $\Delta\beta_f$ and $\Delta\gamma_{\varphi}$ given in
(\ref{betafsconf}).

First, let us work with $V^{(1)}$ on a flat background. Then, choosing
$\mu^2=\varphi_m^2$, where $\varphi_m^2$ is the minimum of the
potential, from the condition
$\displaystyle{ \partial V^{(1)} \over \partial \varphi }=0$ one can
define the
scalar coupling in terms of the remaining ones. Supposing that the QG
coupling $\lambda^2$ is larger than or of the same order as $e^4$
(otherwise, if the $e^4$-term is leading in $\beta_f$ we are in the
situation of Ref. \cite{CW}), we get (for $f \sim e^4+\lambda^2$)
\begin{equation}
V^{(1)}_{\mbox{\scriptsize flat}}={1 \over 48 (4\pi)^2}
\left[
36e^4
+\lambda^2 \xi^2 \left(
15+ {3 \over 4\omega^2} -{9 \xi \over \omega^2}
+{27 \xi^2 \over \omega^2}
\right)
\right]
\varphi^4
\left( \log {\varphi^2 \over \varphi_m^2 } - {1 \over 2} \right) .
\label{V1flat}
\end{equation}
This generalizes the dimensional transmutation mechanism \cite{CW}
so as to include QG modifications. It also shows that {\it
 even at} $e^2=0$
there may exist dimensional transmutation induced by QG. From
(\ref{V1flat}) one can easily find QG corrections for the
scalar-to-vector mass ratio. After shifting the field, the scalar meson
mass is given by
\begin{equation}
m^2(S)=V_{\mbox{\scriptsize flat}}^{(1) \prime \prime}(\varphi_m).
\end{equation}
The photon mass ${m^2(V)=e^2\varphi_m^2}$ appearing after spontaneous
symmetry breaking does not change as compared with the case of no QG.
Thus
\begin{equation}
{ m^2(S) \over m^2(V) }=
{1 \over 6 (4\pi)^2}
\left[
36e^2
+{\lambda^2 \xi^2 \over e^2} \left(
15+ {3 \over 4\omega^2} -{9 \xi \over \omega^2}+{27 \xi^2 \over \omega^2}
\right)
\right] .
\label{m2ratio}
\end{equation}
In the strong gravity regime this is dominated completely
by the second term, i.e. by QG. One may expect that the analog of
(\ref{m2ratio}), where the QG part is universal, may influence the GUT
phenomenology.

We can as well define the relation (\ref{m2ratio}) for the conformal
version of (\ref{Lagr}):
\begin{equation}
{ m^2(S) \over m^2(V) }=
{1 \over 6 (4\pi)^2}
\left(
36e^2
+{5 \over 12}{\lambda^2 \over e^2}
\right).
\end{equation}
QG corrections can again become dominant if $\lambda^2 \gg e^4$. Note
that, taking into account the curvature terms in (\ref{V1}), it is also
possible to define $\xi$ in terms of $e^2, \lambda$
($\xi \sim e^2+\lambda^2$) and find the curvature corrections to
(\ref{m2ratio}).

Now,  we go on to discuss the stability of the EP. It is clear
from expression (\ref{VRGi}) that the RG-improved potential is stable
at large $t$ if $f(t) \ge 0$. Without electromagnetic sector ($e^2=0$)
 and at extremely large $t$, $f(t)$ becomes negative (for the conformal
version) and we observe the above mentioned  QG-driven instability of
the EP.
The same phenomenon takes place in the general version, for reasonable
choices of $\xi$ and $\omega$. However, when the electromagnetic sector
is present, there is always stability for large $t$. (Note that in this
case there exists also a limitation to the $t$-region, as $t$ should be
less than $\displaystyle t_p={3 (4 \pi)^2 \over 2e^2}$, corresponding to
the Landau pole).

In Fig. 1 we show the       behaviour of the scalar coupling
constant for a
case  of the general theory (\ref{Lagr}).
$f(t)$ becomes negative when approaching GUT scales
and leads to instability of the RG-improved
potential for $t < t_i \simeq -6$.  In this $t$-region, the conformal
version does not show any instability.
                                   Note that, since
$\displaystyle {1 \over 2}
\log{\mu_{\mbox{\scriptsize GUT}}^2 \over \mu_{\mbox{\scriptsize Pl}}^2}
\simeq -9$,
the decrease of $t$ from 0 to around --9 corresponds to changing the
energy scale from $\mu_{\mbox{\scriptsize Pl}}^2$ to
$\mu_{\mbox{\scriptsize GUT}}^2$.

We shall now     examine another phenomenon resulting from QG, namely
spontaneous symmetry breaking and phase transitions induced by
curvature.
It is known that in curved space-time spontaneous symmetry breaking
may take place,
in the massless case, already at the classical level if $\xi R>0$,
$\displaystyle\varphi^2={6 \xi R \over f}$.
Taking into account quantum corrections
of matter fields in an external gravitational field, one can find the
possibility of curvature-induced phase transitions (see
\cite{Sh}, for QED in curved space,
a general picture is given in \cite{BOS,BO}). At this point,
we will discuss the same phenomenon for the RG-improved EP in the
presence of QG.

Fig. 2  displays typical shapes of the EP in the
symmetry breaking phase for the conformal version.
This
phase corresponds to positive curvature, and the one   with unbroken
symmetry to $R=0$ ---or also to negative $R$, when taken into
consideration.
In this region (below $\mu_{\mbox{\scriptsize Pl}}$),
the behaviour of
the one-loop potential (for simplicity we keep only the logarithmic
terms in the one-loop corrections) practically coincides with the
behaviour of the RG-improved EP.

It is interesting to estimate the values of the
induced cosmological and Newton couplings in the symmetry breaking
phase. In particular, we may start from $R^2$-gravity with matter
of the form (\ref{Lagr}) without Einstein and cosmological terms, at
energies slightly below the Planck mass. Then, for the choice of
parameters in Fig. 2, we estimate the values of the effective induced
cosmological and Newton constants, which turn out to be the following
(at curvatures $R=10^{-4}$, which is higher than the typical curvature
values in the GUT epoch, i.e. corresponding to the region between
$\mu_{\mbox{\scriptsize GUT}}$ and $\mu_{\mbox{\scriptsize Pl}}$):
\begin{equation}
{1 \over 16\pi G_{\mbox{\scriptsize ind}} }
\simeq 8.2 \cdot 10^{-4} \mu^2,
\hspace{1cm}
{ 2 \Lambda_{\mbox{\scriptsize ind}}
\over 16\pi G_{\mbox{\scriptsize ind}} }
\simeq 4.2 \cdot 10^{-8} \mu^4,
\label{estim}
\end{equation}
where
$\mu_{\mbox{\scriptsize GUT}}< \mu < \mu_{\mbox{\scriptsize Pl}}$.
Similar relations are
true in the conformal version. Such an induced cosmological
constant is too large as compared with experiment, but this may be
reasonably compensated by the ensuing decrease of
 its value due to vacuum energy
contributions at smaller scales.

    Curvature-induced phase transitions  at some
given critical curvature $R_c$ can be found also,
 although they appear for
seemingly small values of $R_c$, corresponding to the range between
the  SM
and GUT scales.
Fig. 3 depicts the behaviour of RG-improved (solid line) versus
one-loop (dashed line) potentials for the general theory.
The first-order phase transition for the RG-improved potential
corresponds to
$R_c \simeq -10^{-9} \mu^2$.

To summarize, we have investigated
in our simple model of scalar QED with quantum $R^2$-gravity the
QG corrections to the effective potential and their
influence on physically measurable phenomena, such as the
vacuum structure, mass ratios, etc.
By virtue of the universality of the QG corrections, our formalism may
certainly be applied to more realistic (and complicated) GUTs.
 Many questions can be studied
in the framework of our formalism for those theories, like the
stability of the scalar sector, the inducement of Einstein gravity,
clearer conections with GUT phenomenology, etc.
In particular, an interesting
possibility is linked with the idea of an inflationary universe (for a
review, see \cite{KT}) where inflation based simply on the
Coleman-Weinberg symmetry breaking mechanism in GUTs is considered to
be unrealistic. It would be of interest to understand how QG corrections
influence the inflationary scenario.

\vskip1cm
\noindent{\large\bf Acknowledgements }

SDO would like to thank I. Antoniadis, M. Einhorn and E. Mottola for
helpful discussions. This investigation has been supported by the
SEP Foundation (Japan), by DGICYT (Spain), and by CIRIT
(Generalitat de Catalunya).

\newpage
\noindent{\large\bf Figure captions}

\noindent{\bf Fig. 1}. Running scalar coupling $f(t)$,
in the general theory with
$\xi=1$, $\omega=1$.
for different initial values of $\lambda(t=0)\equiv\lambda$.
We take $f(t=0)\equiv f$ to be $e^4+\lambda^2$,
where $e\equiv e(t=0)$, for reasons explained in the text.
When setting $\xi={1 \over 6}$, the situation is practically
the same as in the conformal version, and the instability disappears.

\noindent{\bf Fig. 2}. Potential curves for the conformal version.
In this range, the one-loop and RG-improved potentials coincide
to the extent that their associated curves completely overlap
one another. Taking
$\xi=1/6$, $\omega=1$ in the general theory, the curves are
very close to the ones shown.

\noindent{\bf Fig. 3}. One-loop (dashed line) and RG-improved (solid
line) potentials for the general theory, with $\xi=1$, $\omega=1$,
$e^2=10^{-2}$, $\lambda=0.1$, $f=e^4+\lambda^2$, for three different
values of the curvature $R$.
At the scales represented, there is already some quantitative
disagreement between both approaches, particularly in the precise value
of $R_c$, which would be of $\simeq -2.13 \cdot 10^{-9}$ at one loop and
of $\simeq -1 \cdot 10^{-9}$ for the RG-improvement, but the nature of
the overall picture still coincides in both cases.


\end{document}